\documentstyle[epsf,twocolumn,aps]{revtex}
\newcommand{\eq}{\begin{equation}}
\newcommand{\ee}{\end{equation}}
\newcommand{\eqa}{\begin{eqnarray}}
\newcommand{\eea}{\end{eqnarray}}
\def\ab{a_B^*}
\def\lb{l_B}

\def\rxx{\rho_{xx}}

\def\rxy{\rho_{xy}}
\def\bpsi{\bar\psi}
\def\bphi{\bar\phi}
\def\phil{{\phi,l}}
\def\ta{\theta_a}
\def\tb{\theta_b}
\def\eij{\epsilon^{ij}}
\def\br{{\bf r}}
\def\brp{{\bf r}^\prime}
\def\bq{{\bf q}}
\def\bk{{\bf k}}

\def\bkp{{\bf k_+}}
\def\bkm{{\bf k_-}}
\def\ba{{\bf a}}
\def\bb{{\bf b}}
\def\bn{{\bf\nabla}}
\def\rbl{{\bar\rho}_l}
\def\cirr{{\chi_{\rm irr}^0}}
\def\de{\epsilon}
\def\dep{{\epsilon}^\prime}
\def\depp{{\epsilon}^{\prime\prime}}
\def\kinv{{\kappa^{-1}}}
\def\crr{{\chi_{\rho\rho}}}
\def\crrc{{\chi_{\rho\rho}^c}}

\def\dab{D_{\alpha\beta}}
\def\tgamma{{\widetilde\gamma}}
\newcommand{\pprl}{Phys. Rev. Lett. \ } 
\newcommand{\pprb}{Phys. Rev. {B}} 
\begin{document}
\twocolumn[
\hsize\textwidth\columnwidth\hsize\csname@twocolumnfalse\endcsname
\draft
\title{Instabilities Toward Charge Density Wave and Paired Quantum
Hall State of Half-Filled Landau Levels}
\author{Ziqiang Wang}
\address{Department of Physics, Boston College, Chestnut Hill, MA 02167}

\date{\today}
\maketitle

\begin{abstract}

We study the stability of spin-resolved Landau levels at electron
filling factor $\nu=n+1/2$, where $n$ is a positive integer.
Representing the half-filled topmost Landau level by
fermions and the $n$ filled inner Landau levels by $n$ bosons,
coupled to the Chern-Simons gauge fields, we show that
the ground states exhibit charge density wave order for $n(n+1/2)>15.54
(\pi{\ab}^2)n_e$, where $n_e$ is the 2D carrier density and $\ab$ 
is the effective Bohr radius.
We find that the pairing interaction mediated by the fluctuating
gauge field is enhanced near the charge density wave instability
such that p-wave pairing of the Chern-Simons fermions  
prevails for $3.49(\pi{\ab}^2)n_e<n^2(n+1/2)
<20.22(\pi{\ab}^2)n_e$.
The competition between charge density wave order and paired quantum
Hall state is discussed in connection to recent experiments.

\end{abstract}
]

The physics of a sufficiently clean two dimensional electron 
system in strong perpendicular magnetic field has been highlighted by 
its correlated many-body ground state of profound origin known as the 
fractional quantum Hall liquid \cite{books}. 
Recently, it has become evident 
that a different family of correlated ground states emerges
when the Landau level (LL) filling factor is centered around
$\nu=n+1/2$, $n=(0,{\rm positive \ integer})$, corresponding
to $n$ filled inner LLs and a half-filled valence LL.
For $\nu=1/2$, an anomalous ``metal'' state has been known
in which $\rxy$ is not quantized and $\rxx$ shows a shallow minimum
\cite{willett90}. In sharp contrast, quantized Hall plateaus
in $\rxy$ and a low-temperature activation dominated $\rxx$ has
been observed at $\nu=5/2$ \cite{willett87,pan99b},
indicative of an incompressible quantum Hall state. 
Most recently, it has been discovered that beyond the second
LL ($n\ge4$), the low-temperature magnetotransport
becomes highly anisotropic and nonlinear, the strong resistance peak turns
into a deep minimum when the current direction is rotated by 90 degrees
\cite{lilly99a,du99}. The anomalous anisotropic transport can be
interpreted as a manifestation of a spontaneously broken
orientation symmetry in the ground states at $\nu=9/2, 11/2, 13/2,...$
\cite{fradkinkivelson,macdonaldfisher},
such as in a unidirectional charge density wave (CDW) state,
which has been predicted to occur in higher LLs by the Hartree-Fock 
theory \cite{fogler,chalker} and numerical diagonalizations \cite{rezayi99a}.
Experiments have shown
that the isotropic incompressible quantum Hall state at $\nu=5/2$ 
is destroyed when a large in-plane component of the 
magnetic field is present, taking its place is a state with 
highly anisotropic resistances similar to those observed in
higher LLs \cite{lilly99b,pan99}.
That these ground states are close in energy are further
supported by numerical calculations using various interacting potentials 
\cite{rezayi99b,morf}. 

In this paper, we study at zero-temperature the ground state stability
of the half-filled LLs by a generalization of the fermion Chern-Simons 
(CS) theory of Halperin, Lee and Read (HLR) for $\nu=1/2$
\cite{hlr} to $\nu=n+1/2$. In the theory of HLR,
electrons in the lowest LL are represented by fermions bound to two
flux quanta through the introduction of the statistical CS gauge field.
Uniform smearing of the flux in the direction opposite to the
external field leads to a meanfield metallic state in which
the CS fermions experience zero effective magnetic field. Using
this as the starting point, a CS Fermi liquid coupled to strong
gauge field fluctuations has been formulated and shown to correctly
describe the longitudinal responses at $\nu=1/2$.
Naturally, one would like to understand the ways by which such
a Fermi liquid-like state could become unstable due to residual 
quasiparticle interactions mediated by strong gauge field fluctuations.
For example, a p-wave BCS-like pairing instability \cite{pwave} 
of the CS fermions
would indicate the tendency towards the formation of incompressible
paired quantum Hall state \cite{greiter} of the type proposed
by Moore and Read \cite{mooreread}. 
Moreover, a unidirectional CDW instability
would give rise to a state with
spontaneous rotational symmetry breaking and anisotropic transport.

While no evidence has been found for such breakdown of the CS Fermi 
liquid at $\nu=1/2$ ($n=0$), 
we show in this paper that both CDW and p-wave pairing instabilities exist
in higher LLs (finite $n$).
The basic reason for the emergence of these instabilities is
the {\it mutual} screening of interactions between electrons in
the filled inner LLs and in the half-filled valence LL. It produces an
effective interaction between the CS fermions that differs from
the Coulomb interaction at finite $q$ and can become
attractive at short distances.
We calculate the density-density response and show that
the CS Fermi liquid becomes unstable towards a
CDW state for $n\ge 7.77  (\ab/\lb)^2$, {\it i.e.}
$n(n+1/2)>15.54(\pi{\ab}^2)n_e$, where $\ab$ is the effective Bohr radius,
$\lb$ is the magnetic length at $\nu=n+1/2$, and $n_e$ is the
2D carrier density. The onset of the CDW instability occurs
at a wave vector $q_c= 2.41\ab/n\lb^2$. 
Close to the CDW instability, the pair interactions are enhanced by 
the strong fluctuations of the gauge field at finite $q\simeq q_c$. 
We find that the CS Fermi liquid becomes unstable towards a p-wave 
paired (quantum Hall) state for $ 1.32 \ab/\lb <n < 3.18
\ab/\lb$, {\it i.e.} $3.49(\pi{\ab}^2)n_e<n^2(n+1/2)<
20.22(\pi{\ab}^2)n_e$.
Taking $\ab\simeq100$\AA \ for GaAs and
$n_e=2.7-3\times10^{11}$/cm$^2$, these results predict a
CDW instability for $n\ge4$ ($\nu\ge9/2$) and a p-wave paired
quantum Hall state for $n=2$ ($\nu=5/2$) in remarkable agreement
with experimental findings \cite{willett87,pan99b,lilly99a,du99}.

We start by writing down an effective theory at $\nu=n+1/2$,
in which the electrons in the $n$ filled lower LLs 
are represented by $n$ bosons
\cite{wen90} and those in the half-filled valence LL by fermions
\cite{hlr}, coupled to the statistical CS gauge fields.
The physical picture is that the electrons in each of the filled
LL form an independent incompressible state which will be described
as a condensate of the bosons. Excitations above the condensate
will have an energy gap $\hbar\omega_c$. On the other hand, the
electrons in the half-filled topmost LL form a compressible CS Fermi liquid
supporting low-lying quasiparticle excitations and gauge fluctuations.
The Lagrangian density is given by
\eqa
{\cal L}&=&{\cal L}_\psi+{\cal L}_\phi+{\cal L}_I, 
\label{l} \\
{\cal L}_\psi&=& \bpsi(\partial_\tau-ia_0)\psi-{1\over2m}\bpsi(\partial_i
-ia_i+eA_i)^2\psi
\nonumber \\
&-&{i\over2\pi\ta}a_0\eij\partial_i a_j - \mu_f\bpsi\psi, 
\label{la} \\
{\cal L}_\phi&=&\sum_l\bigl[\bphi_l(\partial_\tau-ib_{0,l})\phi_l
-{1\over2m}\bphi_l(\partial_i-ib_{i.l}+eA_i)^2\phi_l
\nonumber \\
&-&{i\over2\pi\tb}b_{0,l}\eij\partial_ib_{j,l}-\mu_l\bphi_l\phi_l\bigr],
\label{lb} \\
{\cal L}_I&=&{1\over2}\int\! d^2\brp
\bigl[\bpsi(\br)\psi(\br)+\sum_l \bphi_l(\br)\phi_l(\br)\bigr]\times
\nonumber \\
&&\times V(\br-\brp)
\bigl[\bpsi(\brp)
\psi(\brp)+\sum_l \bphi_l(\brp)\phi_l(\brp)\bigr].
\label{li}
\eea
Here $\bpsi,\psi$ and $\bphi_l,\phi_l$,
$l=1,\dots n$ are the CS fermion and boson fields respectively,
$m=0.067m_e$ is the band mass in GaAs, and ${\bf A}$ is the vector potential
for the external magnetic field . We work in the Coulomb gauge
where the CS gauge fields $\ba$ and $\bb$ satisfy $\partial_i a_i=
\partial_i b_i=0$.
Integration over $a_0$ and
$b_0$ enforces the constraints on the 2D curl ($\wedge$) of the
gauge fields,
\eqa
\bn\wedge \ba &\equiv& \eij\partial_i a_j=-2\pi\ta\bpsi(r)\psi(r)
\label{curla}\\
\bn\wedge \bb_l &\equiv& \eij\partial_i b_{j,l}=-2\pi\tb\bphi_l(r)\phi_l(r),
\label{curlb}
\eea
which correspond to attaching $\ta$ (even) flux quanta to the fermions
and $\tb$ (odd) flux quanta to each component of the bosons, turning
them back to the original electrons.  For $\ta=2$ and $\tb=1$,
the external magnetic field is canceled out exactly by the vacuum 
expectations of the statistical flux in Eqs ~(\ref{la}) and (\ref{lb}). 
Thus at the mean-field level, both the CS fermions $\psi$, having
a filling fraction $\nu_\psi=1/2$, and the CS bosons $\phi_l$, having
$\nu_\phil=1$, see zero total magnetic field. The electron filling
factor is given by $\nu=\sum_l\nu_\phil+\nu_\psi=n+1/2$.

The electron interaction is described by ${\cal L}_I$ in Eq.~(\ref{li}), where
$V$ is the Coulomb potential, $V(q)=2\pi e^2/\varepsilon q$.
For GaAs, the dielectric constant $\varepsilon\simeq12.6$.
${\cal L}_I$ contains the fermion-fermion and the boson-boson interactions
and, most importantly, the fermion-boson interaction that couples
the topmost and the lower LLs. The latter leads to the {\it mutual}
screening of the interacting potential between the CS fermions by the 
bosons and {\it vice versa}. A natural question arises as to whether one
should treat the fermion or the boson sector of the problem first.
We shall follow a close analogy to the classic electron-phonon
problem in metals, where the answer is known due to Migdal \cite{migdal},
and first solve the boson problem taking into account the screening
by the CS Fermi liquid that has no bosons included. Then
we solve for the screened fermion-fermion interaction using the
renormalized boson properties. In terms of the total symmetric
boson density operator $\rho_s\equiv{1\over\sqrt n}\sum_l\rho_l$,
$\rho_l=\bphi_l\phi_l$, the bosonic part of the interaction becomes,
\eq
{\cal L}_{b,I}={n\over2}\rho_s(\bq,\omega)V_B(q,\omega)
\rho_s(-\bq,-\omega),
\label{lphii}
\ee
with $V_B$ the screened Coulomb interaction,
\eq
V_B(\bq,\omega)=
{2\pi e^2/\varepsilon\over q+ \kappa(\bq,\omega)},\quad
\kappa(\bq,\omega)={2\pi e^2\over\varepsilon} \cirr(\bq,\omega),
\label{vb}
\ee
where $\cirr$ is the fermion (Coulomb) irreducible density-density 
correlation function derived by HLR. In the limit $q<2k_f$ and
$\omega\ll qv_f$,
\eq
\cirr(\bq,\omega)=\left[{2\pi\over m}+{(2\pi\ta)^2\over 24\pi m}
-i(2\pi\ta)^2 {2n_e\over k_f}{\omega\over q^3}\right]^{-1}.
\label{cirr0}
\ee
The screening of the bare Coulomb potential in the lower LLs 
by the compressible CS Fermi liquid
is of the Thomas-Fermi type in Eq.~(\ref{vb}). The screening length
is, after making use of Eq.~(\ref{cirr0}),
$\kappa^{-1}\equiv\kappa^{-1}(q,\omega=0)=\ab(1+\ta^2/12)$ with
the effective Bohr radius $\ab=\varepsilon/me^2$. 
The fact that $V_B$ is short-ranged turns out to be
crucial for modifying the effective fermion
interactions at short distances in the CS Fermi liquid.

With the screened interaction, we now solve the boson part of the problem.
Since the bosons see on average a zero total magnetic field, they
Bose condense into incompressible states, $\langle\phi_l\rangle={\rm const.}$.
The excitations above the condensate can be studied by integrating
out the CS gauge field $\bb_l$ in Eq.~(\ref{lb}) and
decompose the boson field into density and phase fluctuations,
$\phi_l=\sqrt{\rho_l}\exp(i\theta_l)$
and $\rho_l=\rbl+\delta\rho_l$.
Notice that the fermions couple only to the totally symmetric boson
density fluctuations in Eq.~(\ref{li}) which is the only mode affected
by the interaction. We thus perform a unitary rotation
on the $n$ boson fields and keep only the totally symmetric mode:
$(\delta\rho_s,\theta_s)={1\over\sqrt{n}}\sum_l(\delta\rho_l, \theta_l)$.
The resulting effective boson theory is given to quadratic order by,
\eqa
{\cal L}_{\rm eff}^b&=&
{\rbl q^2\over2m}\theta_s(\bq,\omega)\theta_s(-\bq,-\omega)
-i\omega\theta_s(\bq,\omega)\delta\rho_s(-\bq,-\omega)
\nonumber \\
&+&{(2\pi\tb)^2\rbl\over2mq^2}
\delta\rho_s(\bq,\omega)\delta\rho_s(-\bq,-\omega)+{\cal L}_{b,I},
\label{leffb}
\eea
with ${\cal L}_{b,I}$ given in Eq.~(\ref{lphii}). 
The dynamical density-density correlation can be obtained from
Eq.~(\ref{leffb}),
\eq
\langle \delta\rho_s(\bq,\omega)\delta\rho_s(-\bq,-\omega)\rangle
={{(\rbl/ m)}q^2\over \omega_q^2-\omega^2},
\label{rhosrhos}
\ee
where 
$\omega_q^2=\omega_c^2+{(n\rbl/m)}q^2 V_B(\bq,\omega)$,
and the cyclotron frequency $\omega_c=2\pi\tb\rbl/m=eB_{\rm ext}/m$.
This mode corresponds to the cyclotron resonance in accordance with Kohn's
theorem. It is interesting to note that the mode disperses quadratically
due to the screening of Coulomb interaction by the CS Fermi liquid
and becomes overdamped as seen from Eqs~(\ref{vb}) and (\ref{cirr0}).

It is now straightforward to integrate out the boson fluctuations
in Eq.~(\ref{l}), using Eq.~(\ref{rhosrhos}), to obtain an effective
theory for the CS fermions in the topmost LL coupled to gauge field
fluctuations $a_0$ and $\delta\ba=\ba-{\bf A}$,
\eqa
{\cal L}_{\rm eff}^f&=& \bpsi(\bk,\nu)(-i\nu+\varepsilon_k)\psi(\bk,\nu)
+i{qa_0(\bq,\omega)\over 2\pi\ta}a_t(-\bq,-\omega)
\nonumber \\
&+&{1\over 2m}\langle\bpsi\psi\rangle a_t(\bq,\omega)a_t(-\bq,-\omega)
\nonumber \\
&+&\sum_{k,\nu}\sum_{\alpha=0,t}\Lambda^\alpha(\bk,\bq)a_\alpha(\bq,\omega)
\bpsi(\bkp,\omega+\nu)\psi(\bkm,\nu)
\nonumber \\
&+&{1\over2}{1\over(2\pi\ta)^2}a_t(\bq,\omega)q^2{V(q)\over
\de(\bq,\omega)}
a_t(-\bq,-\omega).
\label{lefff}
\eea
Here $\varepsilon_k=k^2/2m-\mu_f$, $\bk_\pm=\bk\pm\bq/2$,
$a_t(\bq)=i\eij(q_i/q)\delta a_j(\bq)$ is the transverse component of
the gauge field, and the fermion-gauge field coupling vertices
\eq
\Lambda^t(\bk,\bq)=-{i\over m}\left({\bk\wedge\bq\over q}\right),\qquad
\Lambda^0(\bk,\bq)=-i.
\label{vertices}
\ee
The renormalized theory differs from the lowest LL
theory of HLR only in the effective fermion-fermion
interaction, the last term in Eq.~(\ref{lefff}), where the dielectric
function governing the mutual screening effects is given by, 
from Eqs~(\ref{li}) and (\ref{rhosrhos}),
\eq
\de(\bq,\omega)=1+{{(n\rbl/ m)}q^2V(q)\over\omega_c^2
+{n\rbl\over m}q^2[V_B(\bq,\omega)-V(q)]-\omega^2},
\label{de}
\ee
with $V_B$ given in Eq.~(\ref{vb}). 
In the limit $\omega/qv_f\ll1$, the dielectric function
behaves as $\de=\dep+i\omega\depp$,
\eqa
\dep(q)&=&{1+\kinv q+\kinv (n\lb^2/\ab)q^2\over 
1-q(n\lb^2/\ab-\kinv)}.
\label{dep} \\
\depp(q) &=&
\left({\ta\over e}\right)^2{
\varepsilon 
k_f(n\lb^2/\ab)^2\over
[1-q(n\lb^2/\ab-\kinv)]^2}.
\label{depp}
\eea
As a result of the mutual screening, the effective interaction between
the CS fermions is significantly modified at finite $q$, and becomes
attractive at large enough $q$ so long as $\kappa> \ab/n\lb^2$. Notice that
one recovers to leading order the dielectric function obtained by 
Aleiner and Glazman \cite{aleiner95} from projecting out the lower LLs, 
which was used in the Hartree-Fock studies \cite{fogler},
in the limit $\kappa\to0$, {\it i.e.} neglecting the Thomas-Fermi
screening of the lower LL interaction by the compressible CS Fermi liquid in
the topmost LL. We next demonstrate that the effective theory given
in Eqs~(\ref{lefff})-(\ref{depp}) produces CDW and pairing instabilities of 
the CS Fermi liquid outside the regime of the Hartree-Fock theory.

From Eq.~(\ref{curla}), it follows that the fermion density-density
correlation function is given by
\eq
\crr(\bq,\omega)=(2\pi\ta)^{-2}q^2\langle a_t(\bq,\omega)
a_t(-\bq,-\omega)\rangle.
\label{crr}
\ee
To determine the gauge field propagators, $\dab(\bq,\omega)\equiv
\langle a_\alpha(\bq,\omega)a_\beta(-\bq,-\omega)\rangle$, we follow the
RPA approach of HLR and integrate out the
fermion fields in Eq.~(\ref{lefff}) to arrive at an effective action
for the gauge field,
\eq
S_{\rm eff}={1\over2}\sum_{\bq,\omega}\sum_{\alpha\beta}
a_\alpha(\bq,\omega)\dab^{-1}(\bq,\omega)
a_\beta(-\bq,-\omega),
\label{seff}
\ee
where the inverse propagator matrix is given by
\eq
D^{-1}(\bq,\omega)=\left(\begin{array}{cc}
{m/2\pi} & i{q/2\pi\ta} \\
i{q/2\pi\ta} & \chi(\bq,\omega)q^2-i\gamma\omega/q 
\end{array}\right),
\label{dab}
\ee
for $q<2k_f$ and $\omega\ll  q v_f$. In Eq.~(\ref{dab})
$\gamma=2n_\psi/k_f$ and
$\chi(\bq,\omega)=(1/24\pi m)+V(q)/\de(\bq,\omega)(2\pi\ta)^2$.
The density response is then, from Eqs (\ref{crr}) and (\ref{dab}),
\eq
\crr(\bq,\omega)={\varepsilon/2\pi e^2
\over \kinv+[1/q\dep(q)]-i\varepsilon\tgamma(q)\omega/2\pi e^2q^3},
\label{crr1}
\ee
where $\tgamma(q)=(2\pi\ta)^2\gamma+q^2\depp(q)/[\dep(q)]^2$.
Substituting the dielectric functions in Eqs~(\ref{dep})
and (\ref{depp}) into Eq.~(\ref{crr1}), one finds that $\crr(\bq,0)$
may diverge at a finite $q=q_c$, giving rise to CDW formation.
Near $q_c$, we obtain,
\eq
\crrc(\bq,\omega)={A\over (q-q_c)^2+\Delta-iA\tgamma(q_c)\omega/q_c^3},
\label{crrc}
\ee
where $A$ is a $q_c$-dependent constant and
\eq
\Delta=\kappa{\ab\over n\lb^2}-{1\over4}\bigl(\kappa-{\ab\over n\lb^2}\bigr)^2,
\quad q_c={1\over2}\bigl(\kappa-{\ab\over n\lb^2}\bigr).
\label{deltac}
\ee
The onset of the instability, {\it i.e.} the critical 
point for the zero temperature phase transition, is marked by $\Delta=0$,
close to which $A=1.71q_c^3\varepsilon/2\pi e^2$.
When $\Delta<0$, which happens for $n> (3+2\sqrt{2})(1+\ta^2/12)(\ab/\lb)^2$,
the Fermi liquid-like ground state becomes unstable and is replaced
by a CDW state with an initial ordering wave vector 
$q_c=2.414\ab/n\lb^2$.
Since the charge density is related to the statistical
flux density through Eq.~(\ref{curla}), the system becomes energetically
more favorable to spontaneously condensing the statistical flux into
a flux density wave rather than uniformly smearing the gauge flux
which is the starting point for the CS Fermi liquid, making the
latter unstable. One expects, based on energetic considerations of
this effect when there is particle-hole symmetry in the half-filled topmost 
LL, that the ordering vector $\bq_c=(0,\pm q_c), (\pm q_c,0)$.
As a result of the enhanced (divergent) static susceptibility near the
instability, a unidirectional CDW state in the absence of an in-plane field
can be induced by a weak sample anisotropy which may be born out of the 
underlying crystallography conditions.
For $\ta=2$, {\it i.e.} half-filled valence LL, using $n_e=\nu/2\pi\lb^2$,
we can rewrite the CDW instability condition as $n(n+1/2)>15.54
(\pi{\ab}^2)n_e$. Taking $\ab\sim100$\AA \ for GaAs and a 2D carrier
density $n_e=2.7-3 \times 10^{11}/$cm$^2$, the CDW order
emerges for $n\ge4$, {\it i.e.} for $\nu\ge9/2$, consistent with
experimental findings \cite{lilly99a,du99}.

We next turn to the pairing instability of the CS Fermi liquid towards
the paired quantum Hall state. The proximity
to the CDW formation enhances the pair interactions mediated by the
gauge field fluctuations. The static pair interaction in the Cooper channel
is given by
\eq
\Gamma(\bkp,\bkm)
=-\sum_{\alpha\beta}\Lambda^\alpha(\bk,-\bq)D_{\alpha\beta}(\bq)
\Lambda^{\beta}(-\bk,\bq),
\label{gamma}
\ee
where the coupling vertices $\Lambda$ are given in Eq.~(\ref{vertices}).
In the long wavelength limit, the attractive interaction mediated
by the current-density fluctuations ($D_{0t}$) in the
p-wave channel \cite{greiter} is known to be overcome by the logarithmically
singular pair-breaking interaction mediated by the transverse
current-current fluctuations \cite{bonesteel}. However, close to the
CDW instability, gauge fluctuations in all channels are
significantly enhances near $q_c$,
\eq
D(\bq)={A\over (q-q_c)^2+\Delta}\left[\begin{array}{cc}
-({2\pi\over m})^2 & -i{2\pi\over m}{2\pi\ta\over q_c} \\
-i{2\pi\over m}{2\pi\ta\over q_c} & ({2\pi\ta\over q_c})^2
\end{array}\right],
\label{d}
\ee
where $\Delta$ and $q_c$ are given in Eq.~(\ref{deltac}),
and become singular as $\Delta\to0$. These contributions 
dominate over those in the small $q$ limit and
allows the attractive part of the interactions to compete, resulting in a
pairing instability. Within the standard BCS framework, we determine
the pairing instability by calculating the effective interaction
averaged on the Fermi surface in the $l$-angular momentum channel,
\eq
\lambda_l=-{m\over(2\pi)^2}\int d\varphi_+ d\varphi_-
\Gamma(\bkp,\bkm)e^{i(\varphi_+-\varphi_-)l},
\label{lambdal}
\ee
where $\varphi_\pm$ are the angles 
that $\bkp$ and $\bkm$ make with the $x$-axis on the Fermi circle:
$\vert\bkp\vert^2=\vert\bkm\vert^2=k_f^2$. The sign convention is
such that $\lambda_l>0$ corresponds to attractive interactions in
the $l$-wave channel. For our spin-polarized case, pairing is only allowed
for $l={\rm odd}$ integer. In the $l=1$ p-wave channel,
we calculate $\lambda_1$ using Eqs~(\ref{gamma})-(\ref{lambdal}) near
the CDW instability, {\it i.e.} for $0<\Delta\ll q_c^2$,
\eq
\lambda_1\simeq{2\pi^2A\over m k_f\sqrt{\Delta}}
\bigl({2k_f\over q_c}\bigr)^2
\biggl[-1+8\bigl({q_c\over2k_f}\bigr)^2-8\bigl({q_c\over2k_f}\bigr)^4
\biggr],
\label{lambda1}
\ee
where we have put $\ta=2$. The condition for $\lambda_1>0$
is thus given by
$0.38<q_c/2k_f<0.92$ or equivalently $1.32\ab/\lb < n <3.18\ab/\lb$.
As in the CDW case, we can solve these inequalities to obtain,
$3.49(\pi{\ab}^2)n_e < (n+1/2)n^2 <20.22(\pi {\ab}^2)n_e$
as the condition for the p-wave pairing instability of the CS Fermi liquid
and the emergence of the p-wave paired quantum Hall states.
Using the $\ab$ for GaAs and $n_e=2.3-3 \times 10^{11}/$cm$^2$,
we find that the paired state is possible within the
RPA for $n=2$, {\it i.e.} $\nu=5/2$ in good agreement with 
experimental observations 
\cite{willett87,pan99b}.

We have concentrated in this paper on half-filling LLs which
correspond to choosing $\ta=2$. The same theory applies to cases when the
electron filling factor in the topmost LL is at other even denominator
fractions, {\it e.g.} when $\nu=n+1/2k$ and $\ta=2k$. 
Eqs~(\ref{crr})-(\ref{deltac}) show that the CDW instability also exists
in these flanks of the LL where re-entrant integer quantum Hall
states have been observed experimentally \cite{lilly99b}.

The author thanks D. Broido, J. Engelbrecht, Y.~B. Kim, and N. Nagaosa 
for useful discussions. This work was supported in part by DOE Grant No. 
DE-FG02-99ER45747 and an award from Research Corporation.

\end{document}